\documentclass[12pt]{article}
\pdfoutput=1
\usepackage[utf8]{inputenc}
\usepackage{amsmath}
\usepackage{amssymb}
\usepackage{graphicx}
\usepackage{slashed}
\usepackage{array}
\usepackage[leftcaption,raggedright]{sidecap}
\usepackage{caption}
\usepackage[hidelinks]{hyperref}
\usepackage{siunitx}

\newcommand\pubnumber{NuPhys2026-Adam-Lister}
\newcommand\pubdate{\today}

\def\uwmadison{University of Wisconsin - Madison}

\def\support{\footnote{
  This document was prepared by The NOvA Collaboration using the resources of the Fermi National Accelerator Laboratory (Fermilab), a U.S. Department of Energy, Office of Sicence, Office of High Energy Physics HEP User Facility. We would like to thank the MiniBooNE collaboration for providing their flux simulation for the Booster Neutrino Beam.
}}

\def\Title#1{\begin{center} {\Large #1 } \end{center}}
\def\Author#1{\begin{center}{ \sc #1} \end{center}}
\def\Address#1{\begin{center}{ \it #1} \end{center}}

\newcommand\pubblock{\rightline{\begin{tabular}{l} \pubnumber\\
         \pubdate  \end{tabular}}}
\newenvironment{Abstract}{\begin{quotation}  }{\end{quotation}}
\newenvironment{Presented}{\begin{quotation} \begin{center} 
             PRESENTED AT\end{center}\bigskip 
      \begin{center}\begin{large}}{\end{large}\end{center} \end{quotation}}

\def\beq{\begin{equation}}
\def\eeq#1{\label{#1}\end{equation}}
\def\eeqn{\end{equation}}

\def\beqa{\begin{eqnarray}}
\def\eeqa#1{\label{#1}\end{eqnarray}}
\def\eeqan{\end{eqnarray}}

\let\bar=\overbar

\def\Dslash{\not{\hbox{\kern-4pt $D$}}}
\def\dslash{\not{\hbox{\kern-2pt $\del$}}}

\def\msb{{\bar{\ssstyle M \kern -1pt S}}}

\begin{document}
\begin{titlepage}
\pubblock

\vfill
\Title{NOvA's Current and Future Sterile Neutrino Searches}
\vfill
\Author{ Adam Lister\support}
\Address{\uwmadison}
\vfill
\begin{Abstract}
The NOvA experiment’s most recent search for eV-scale sterile neutrinos under a 3+1 model simultaneously analyses muon neutrino and neutral current datasets from the NuMI beam at its Near ($\sim$\qty{1}{km} baseline) and Far (\qty{810}{km} baseline) detectors to look for oscillations consistent with a sterile neutrino. The analysis is systematically limited in the region of parameter space where $\Delta m^2_{41} \gtrsim 1~\mathrm{eV}^2$. This region of parameter space is preferred by sterile neutrino interpretations of current experimental anomalies and so improving sensitivity here is high-priority. These proceedings present our current search strategy, and discusses future plans to include data from a second beamline, the Booster Neutrino Beam, to improve our sensitivity in systematics-dominated regions of parameter space.
\end{Abstract}
\vfill
\begin{Presented}
NuPhys2026, Prospects in Neutrino Physics\\
King's College, London, UK,\\ January 7--9, 2026
\end{Presented}
\vfill
\end{titlepage}
\def\thefootnote{\fnsymbol{footnote}}
\setcounter{footnote}{0}

\section{Introduction}

The NOvA Experiment is a long-baseline neutrino oscillation experiment with two functionally identical detectors \qty{14}{\milli\radian} off-axis of the NuMI beamline. The Near Detector (ND) is located on-site at Fermilab, Illinois, USA, around \qty{1}{km} downstream of the beam target, and \qty{100}{m} underground, while the Far Detector (FD) is on the surface \qty{810}{km} away, in Ash River, Minnesota, USA. The experiment recently celebrated 10 years of data taking, and has published the world's most precise single-experiment measurement of the atmospheric mass splitting \cite{x53y-2b86}. NOvA has a broad physics programme beyond three-flavour oscillations, covering interaction measurements, searches for new physics in neutrino oscillations, and searches for exotic phenomena more broadly.

Neutrino mixing is now a well established phenomenon, having been measured by numerous experiments over several decades. This process requires that neutrinos have mass, and that those masses not be degenerate. How neutrinos get their masses is an open area of research. Many proposed mass-generation mechanisms are built on the assumption that at least one additional right-handed neutrino exists. This neutrino would not interact with the weak force, making it effectively invisible, though it could impact neutrino oscillations, making oscillations a unique window into this physics.

The global picture on the existence of sterile neutrinos is far from clear. While several experiments have reported anomalous results that could be interpreted as oscillations governed by a new eV-scale neutrino~\cite{PhysRevD.103.052002,PhysRevLett.128.232501,PhysRevLett.133.201804}, many experiments have searched and found no evidence for these particles~\cite{microboone-nature,katrin-nature}. Taken together, the global data has significant tension when interpreted under a ``3+1'' model with one sterile neutrino in addition to the known three. The 3+1 model continues to have a place in neutrino physics as a way for experiments to compare results while searching for any new physics that has some oscillatory behaviour. Additional data in this sector is extremely valuable as we investigate models that can describe the constellation of existing results. 

\section{Sterile Neutrino Searches at NOvA}

\begin{figure}[!t]
  \includegraphics[width=0.49\linewidth]{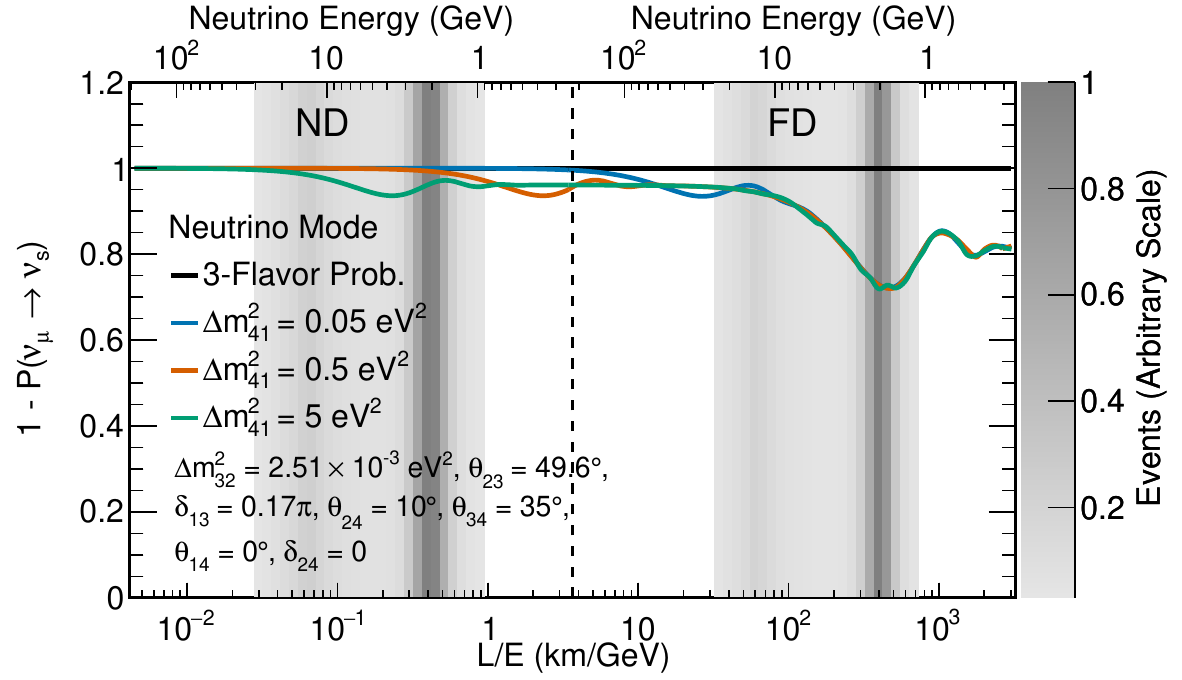}
  \includegraphics[width=0.49\linewidth]{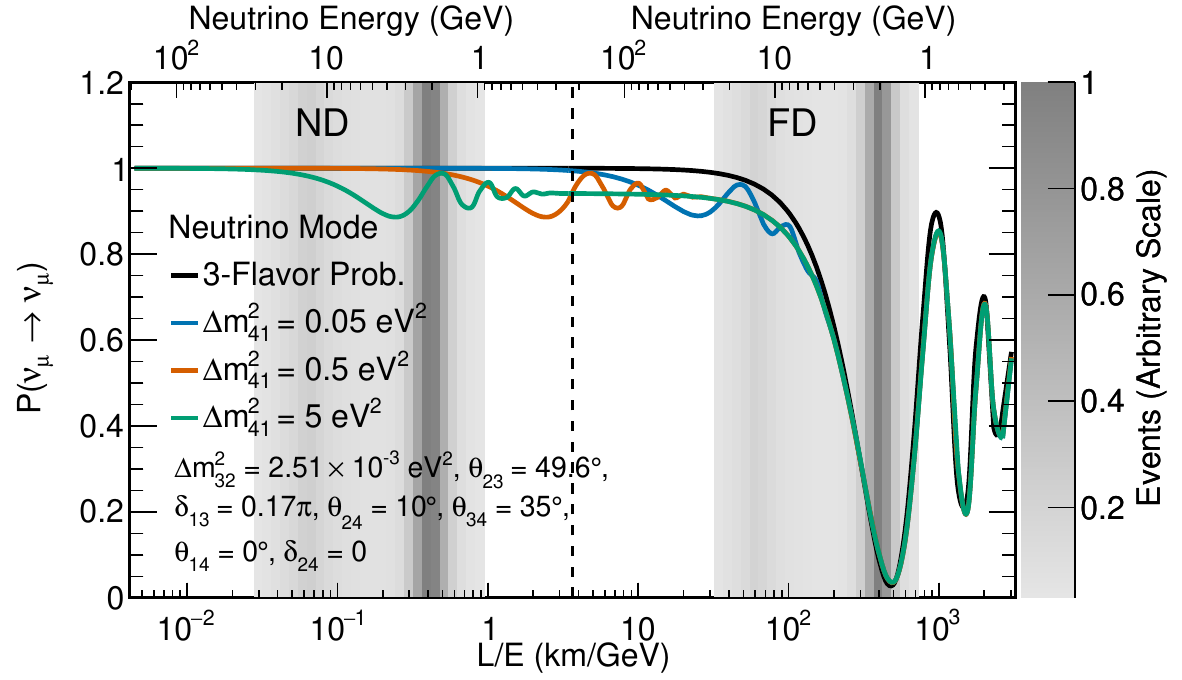}
  \caption{Oscillation probabilities as a function of $L/E$ for NC (left) and muon neutrino (right) disappearance under three-flavour oscillations (black) and with oscillations under a 3+1 model with varying values of $\Delta m^2_{41}$. The shaded grey background corresponds to where the NOvA data lay in the ND, at low $L/E$ and in the FD, at higher $L/E$.}
  \label{fig:oscillograms}
\end{figure}

Historically, NOvA has published searches for sterile neutrinos by looking for disappearance of Neutral Current (NC) interactions at the NOvA FD \cite{PhysRevD.96.072006,PhysRevLett.127.201801}, where the approximate NC survival probability under the 3+1 model is~\cite{MINOS:2016viw}
\begin{equation}\label{eq:prob_nc}
    \begin{aligned}
    1 - P(\nu_{\mu} \rightarrow \nu_{s}) \approx 1 & - \cos^{4}\theta_{14}\cos^{2}\theta_{34}\sin^{2}2\theta_{24}\sin^{2}\Delta_{41} \\
    & - \sin^{2}\theta_{34}\sin^{2}2\theta_{23}\sin^{2}\Delta_{31} \\
    & + \frac{1}{2}\sin\delta_{24}\sin\theta_{24}\sin2\theta_{23}\sin\Delta_{31},
    \end{aligned}
\end{equation}
where $\theta_{14}$, $\theta_{24}$, and $\theta_{34}$ are new mixing angles introduced by the 3+1 model, $\delta_{24}$ is a new phase, and $\Delta_{ji} \equiv \frac{\Delta m_{ji}^{2} L}{4E_{\nu}}$, where $L$ and $E_{\nu}$ are the neutrino baseline and energy. This is a very powerful technique: the active neutrino flavours each participate in NC interactions at the same rate, but sterile neutrinos cannot undergo weak interactions. Any reduction in NC interactions compared to prediction could therefore be indicative of a sterile neutrino (Fig.~\ref{fig:oscillograms}, left). This technique is limited, however, as searching for oscillations only in the FD restricts sensitivity to small values of the new mass splitting, far from current anomalies. The NC channel is also challenging, as the outgoing lepton is a neutrino, which doesn't deposit energy in the detector, and so reconstruction of the neutrino energy becomes biased and more smeared, reducing sensitivity to oscillations.

\begin{figure}[!b]
  \includegraphics[width=0.32\linewidth]{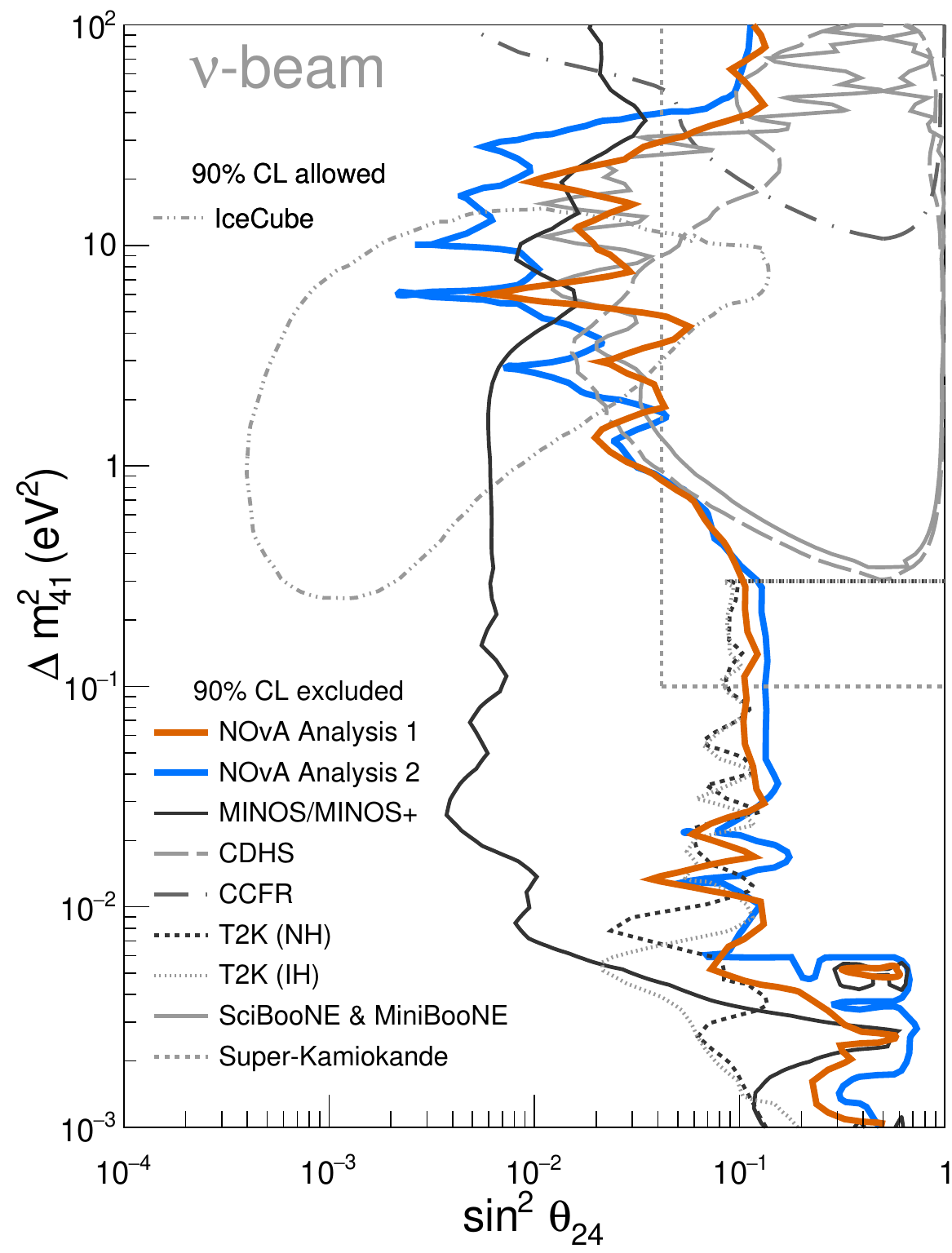}
  \includegraphics[width=0.32\linewidth]{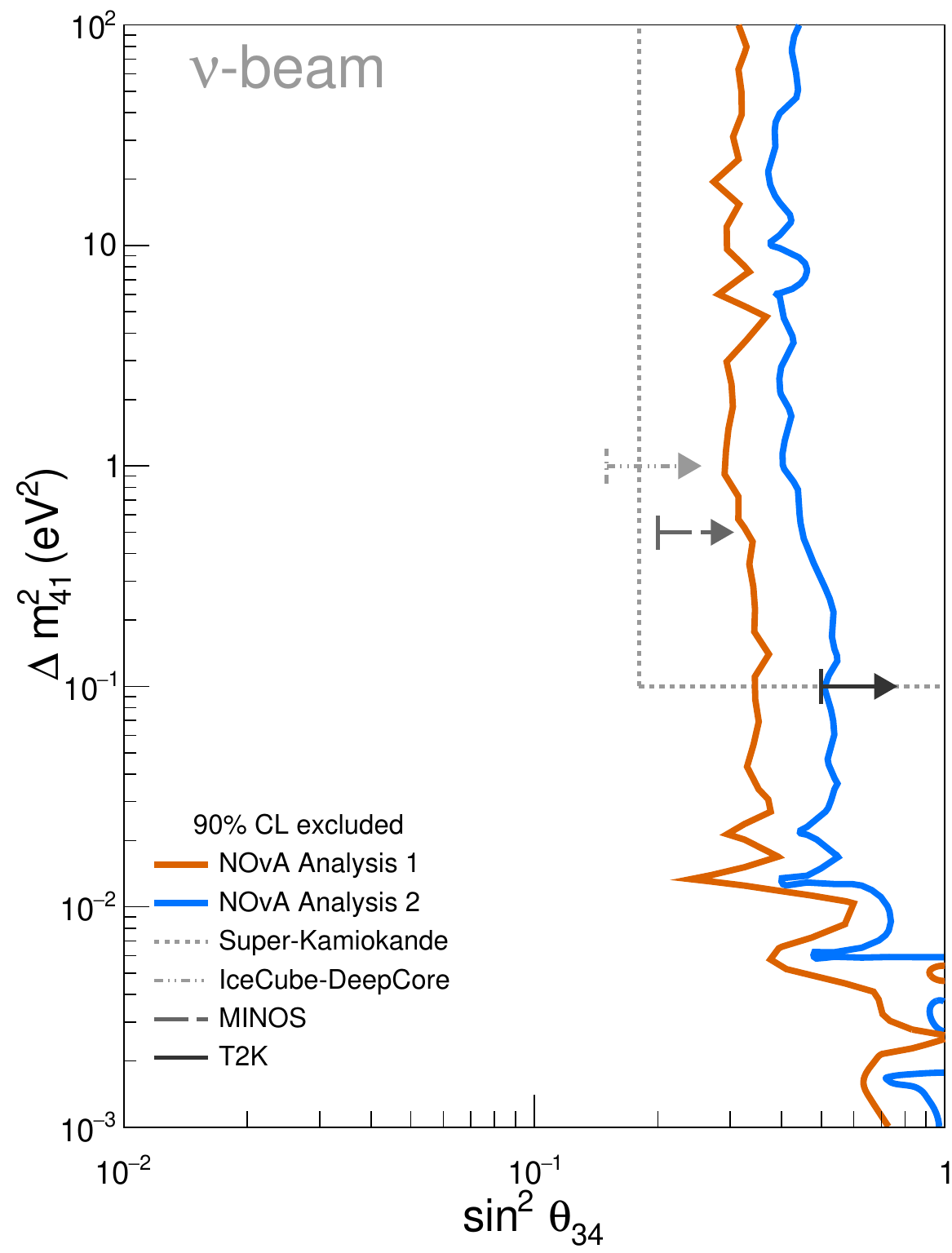}
  \includegraphics[width=0.32\linewidth]{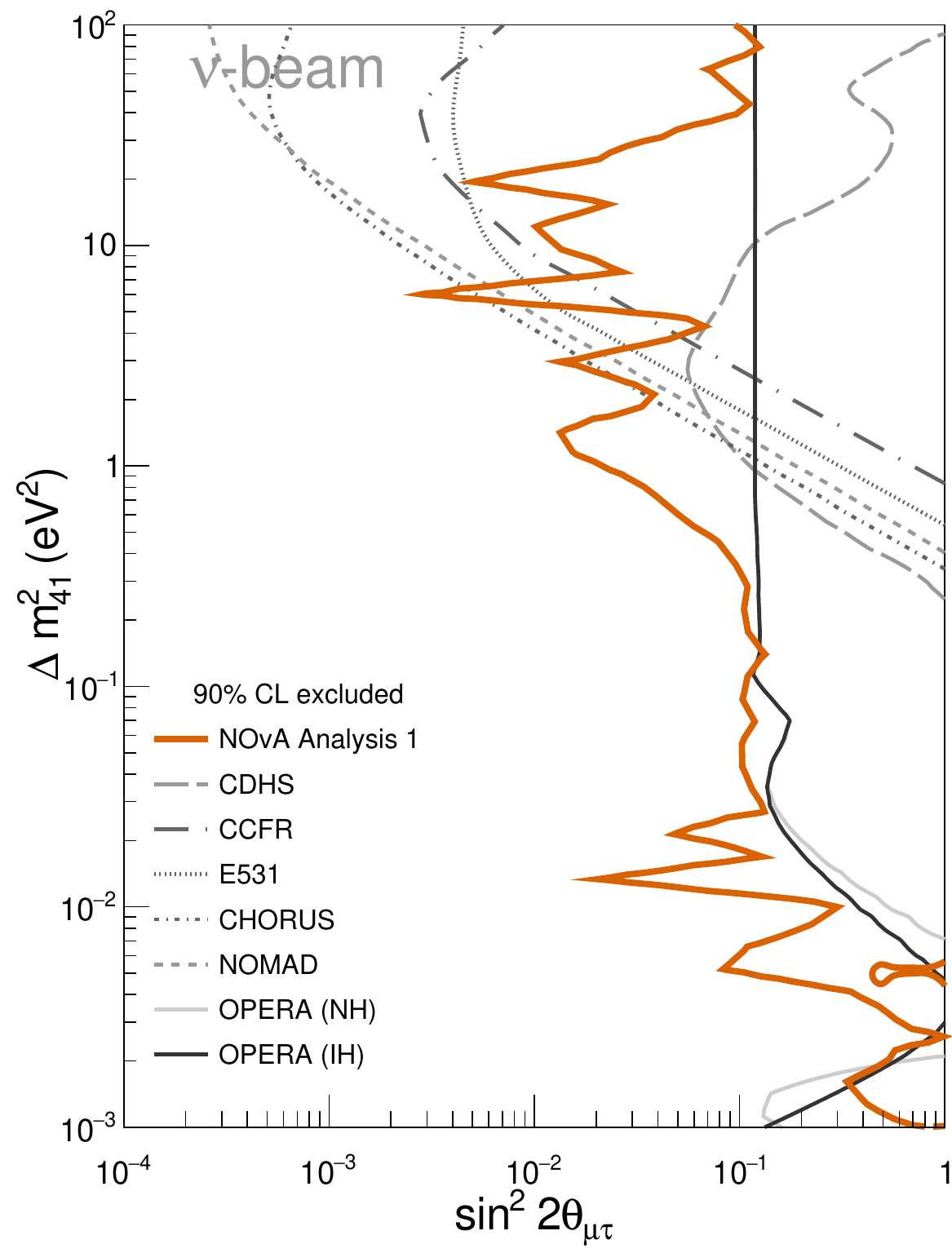}
  \caption{90\% confidence level exclusion limits for NOvA's 2025 sterile neutrino analysis \cite{PhysRevLett.134.081804}, presented in terms of $\Delta m^2_{41}$, and $\sin^2\theta_{24}$, $\sin^2\theta_{34}$, and $\sin^2\theta_{\mu\tau} = \sin^22\theta_{24}\sin^2\theta_{34}$, and compared with results from other experiments. The limits are world-leading in several regions of parameter space.}
  \label{fig:2025sterileresults}
\end{figure}

For our most recent analysis, we used a new technique, simultaneously searching for oscillations in both the ND and FD data, which allows us to place limits across a much broader range of mass-squared splittings~\cite{PhysRevLett.134.081804}. We additionally expanded the search to use the muon neutrino charged-current ($\nu_{\mu}$ CC) disappearance channel in addition to NC disappearance. The $\nu_{\mu}$ survival probability can be approximated as
\begin{equation}\label{eq:prob_cc_numu}
    \begin{aligned}
    P(\nu_{\mu} \rightarrow \nu_{\mu}) \approx 1 & - \sin^{2}2\theta_{24}\sin^{2}\Delta_{41} \\
    & + 2\sin^{2}2\theta_{23}\sin^{2}\theta_{24}\sin^{2}\Delta_{31} \\
    & - \sin^{2}2\theta_{23}\sin^{2}\Delta_{31}.
    \end{aligned}
\end{equation}

The $\nu_{\mu}$ CC channel sees standard three-flavour oscillations in the FD, as visible in the last term of Eq.~\ref{eq:prob_cc_numu}. Here, the presence of a sterile neutrino would provide some additional source of disappearance (Fig.~\ref{fig:oscillograms}, right), which could manifest with any frequency. 

Fitting the four samples (ND $\nu_{\mu}$ CC, ND NC, FD $\nu_{\mu}$ CC, FD NC) simultaneously, we find that our data is consistent with three-flavour oscillations at the 90\% confidence level. We place limits on the sterile neutrino mixing angles $\sin^2\theta_{24}$, $\sin^2\theta_{34}$, and the effective mixing angle $\sin^2\theta_{\mu\tau}$ as a function of $\Delta m^2_{41}$ for the range $10^{-3}$ eV$^2$ $< \Delta m^2_{41} <$ 100 eV$^2$ (Fig.~\ref{fig:2025sterileresults}). These limits are world leading in various regions of parameter space. 

At low values of $\Delta m^2_{41}$, and for the entire $\sin^2\theta_{34}$ space, our sensitivity comes primarily from FD samples, where we are statistically limited, and so additional data is the primary way in which we will be able to improve our sensitivity. We are currently working on a follow-up analysis to that presented in reference \cite{PhysRevLett.134.081804}, which will include twice the neutrino-mode data, our antineutrino-mode data, and a ND $\nu$-on-$e$ sample, which will act as an in-situ flux constraint.

As we move up to larger $\Delta m^2_{41}$ values, oscillations become much more rapid, taking place between the beam target and the ND, and so our sensitivity comes primarily from the ND samples, with oscillations becoming averaged out at the FD. These samples contain hundreds of thousands to millions of events, meaning we are systematics limited in this region of parameter space. The next analysis, with its antineutrino-mode data, will help to constrain some of these uncertainties in this sytematics-limited region. To increase sensitivity beyond this expected improvement, we need to be more creative.

\section{The Booster Neutrino Beam}

\begin{figure}[!t]
  \centering
  \includegraphics[width=1.0\linewidth]{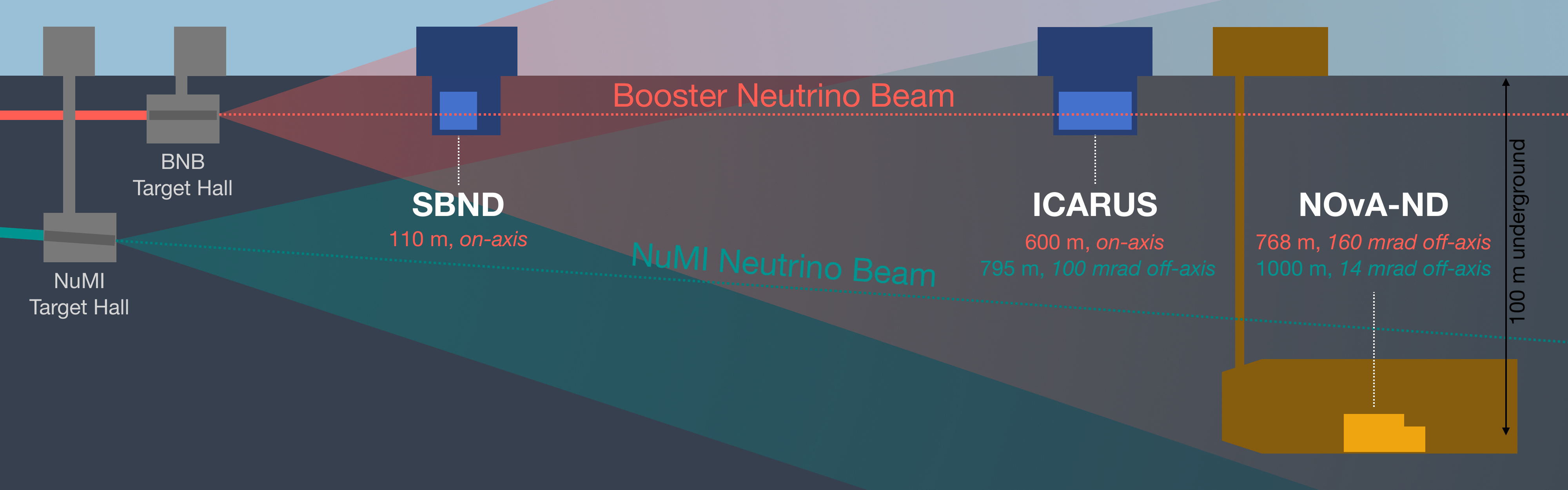}
  \caption{Schematic showing the layout of the two neutrino beams provided by Fermilab, the BNB and NuMI, along with the three currently operating neutrino oscillation experiments. The NOvA ND can see neutrinos from the BNB at an off-axis angle of around \qty{160}{mrad} (\qty{9.2}{\degree}).}
  \label{fig:experiments_schematic}
\end{figure}

Neutrino beams have significant dispersion, meaning that they can be observed off-axis. The NOvA ND's position on-site at Fermilab means that it is able to see neutrinos from the second neutrino beam produced at Fermilab, the Booster Neutrino Beam (BNB), in addition to those from NuMI (Fig.~\ref{fig:experiments_schematic}). The NOvA ND has been taking data \qty{9.2}{\degree} off-axis of the BNB, and $\sim$\qty{770}{m} from the BNB target, using a dedicated trigger since 2015, though it is yet to be used for analysis. During this time, the BNB has exclusively run in neutrino-mode running. This beam is of particular interest in sterile neutrino searches as it is the  beam that the MiniBooNE experiment observed anomalous excesses of electron neutrinos and antineutrinos~\cite{PhysRevD.103.052002}. The SBND, ICARUS, and MicroBooNE experiments are also situated on-axis of the BNB, and are also searching for sterile neutrinos.

\begin{figure}[!b]
  \centering
  \includegraphics[width=0.6\linewidth]{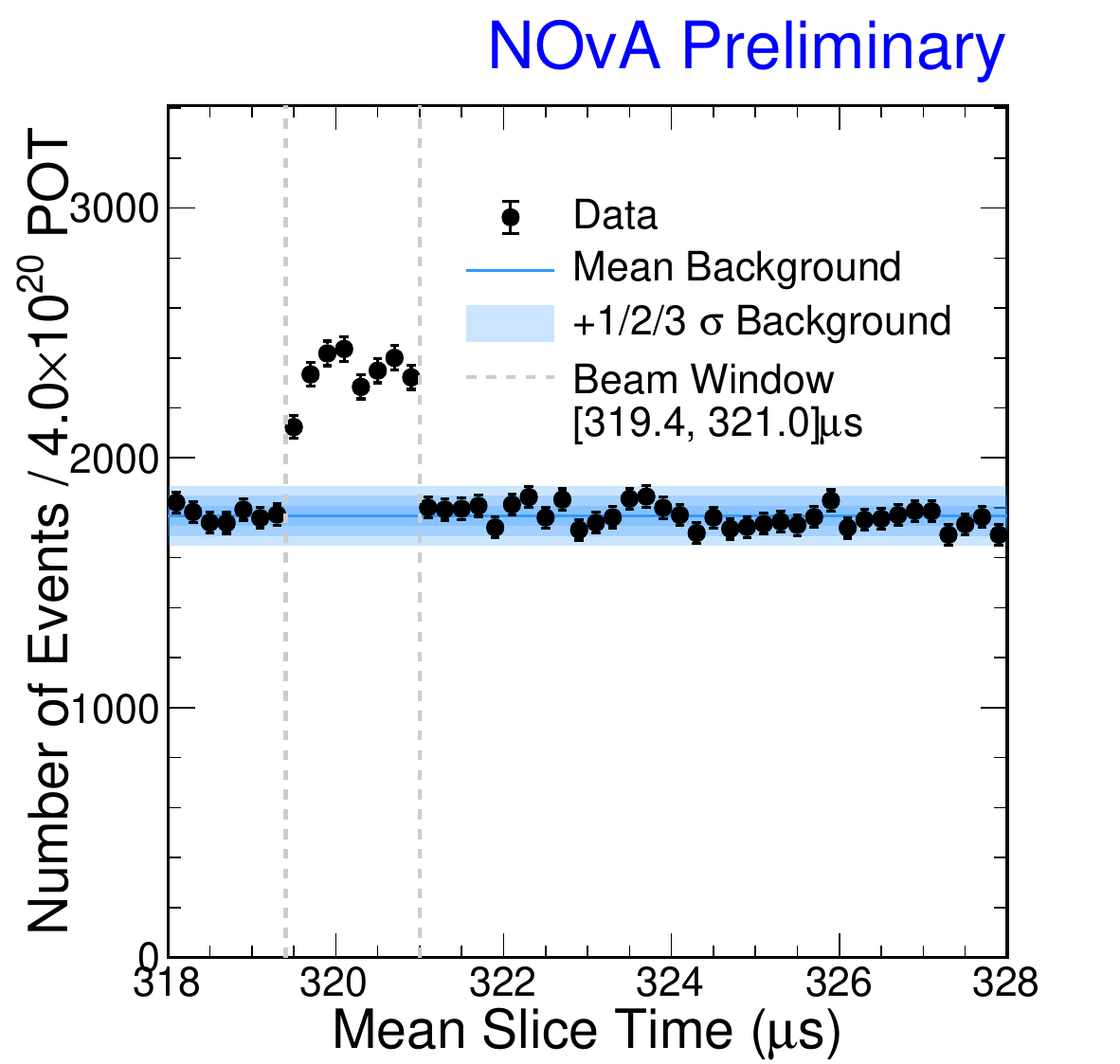}
  \caption{Plot of mean time of each reconstructed ``slice'' in BNB triggered NOvA readout using a subset of the data taken. The peak beginning at \qty{319.4}{\micro\second} corresponds to the expected BNB arrival time, with the width of the distribution corresponding to the expected \qty{1.6}{\micro\second} beam arrival time distribution. Backgrounds are from cosmic rays.}
  \label{fig:top_hat}
\end{figure}

To confirm that the data we have taken contains BNB neutrinos, we can plot the number of events in the BNB triggered data as a function of the time of each reconstructed ``slice'' in the event, as is shown in Fig.~\ref{fig:top_hat}. Here we see a constant backgound which comes from cosmic rays coincident with the BNB trigger, and a peak beginning around \qty{319.4}{\micro\second} which corresponds to the beam arrival time within the readout window for the BNB, which lasts for the expected time of \qty{1.6}{\micro\second} before returning to baseline.

\begin{figure}[!b]
  \centering
  \includegraphics[width=0.49\linewidth]{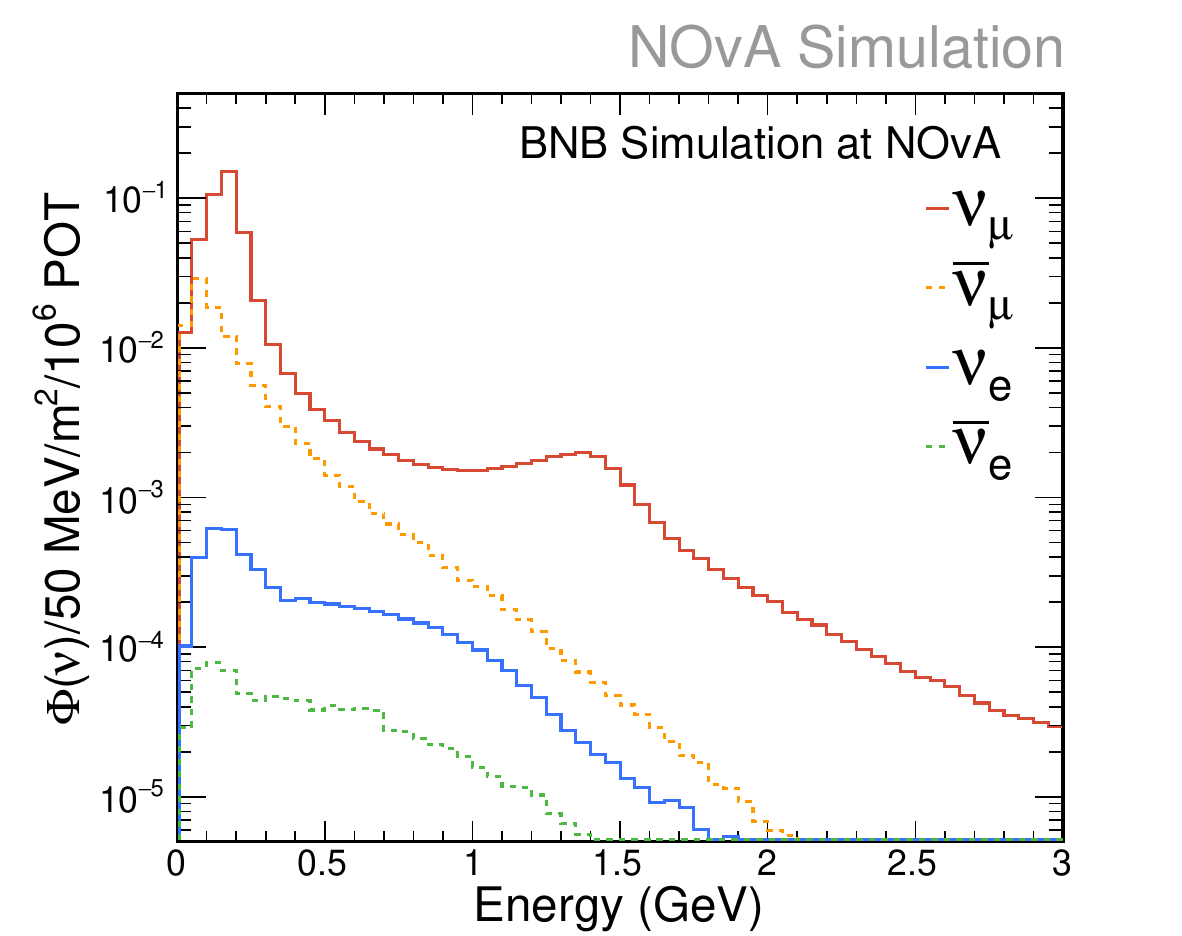}
  \includegraphics[width=0.49\linewidth]{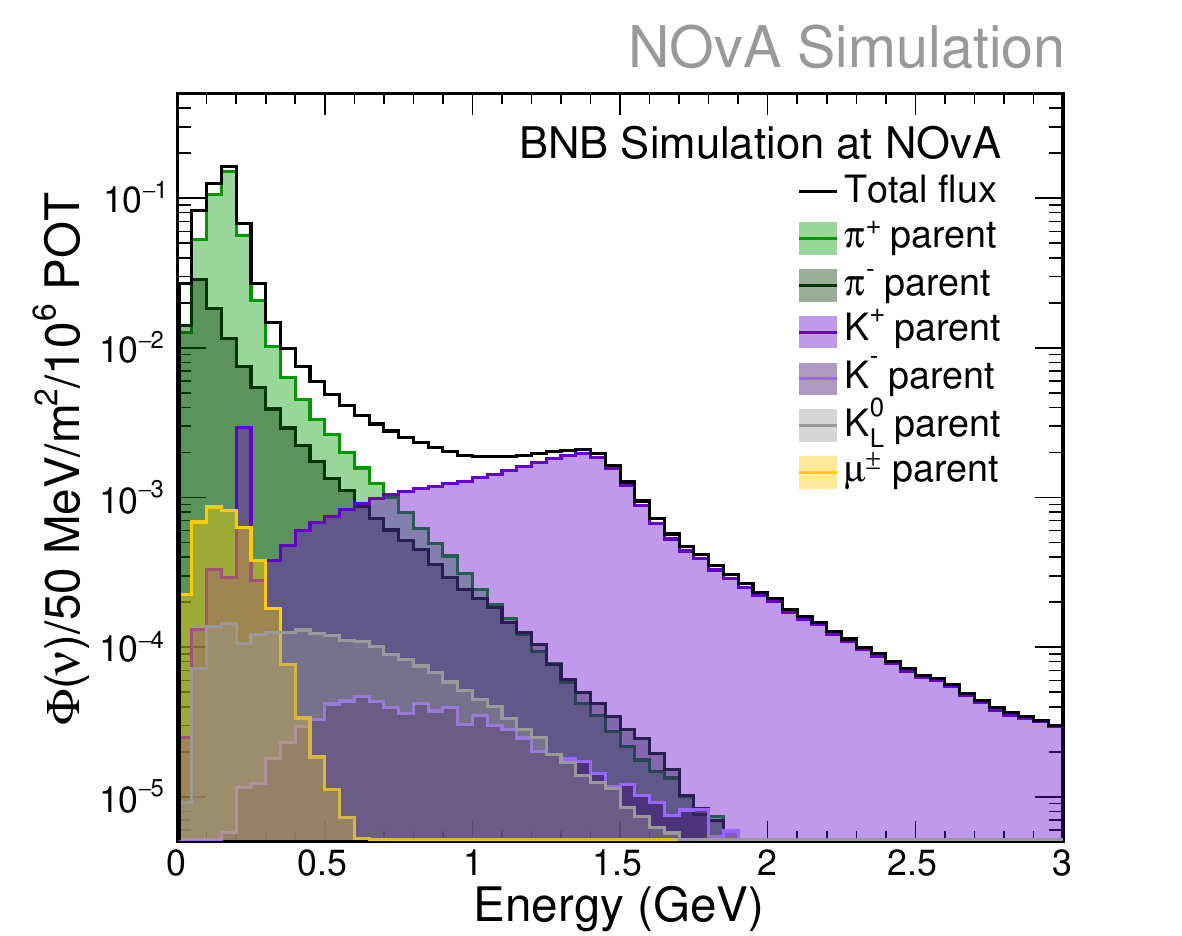}
  \caption{BNB neutrino flux at NOvA ND, separated by neutrino flavour (left), neutrino parent (right). The beam is primarily muon neutrinos, with subdominant contributions from muon neutrinos, electron neutrinos and electron antineutrinos. The beam is dual-peaked with the low-energy peak coming from neutrinos produced in $\pi$ decay, and the high-energy peak coming from neutrinos produced in $K$ decay.}
  \label{fig:flux_prediction}
\end{figure}

Using the flux simulation initially developed by the MiniBooNE collaboration~\cite{PhysRevD.79.072002}, we have generated a prediction for the BNB neutrino flux at the NOvA ND (Fig.~\ref{fig:flux_prediction}). The flux is primarily composed of muon neutrinos, with subdominant contributions from muon antineutrinos and electron neutrinos and antineutrinos. The flux distribution is dual-peaked, with one peak at around 200 MeV, coming from neutrinos with $\pi^+$/$\pi^-$ parents, and another at around 1.4 GeV, coming from neutrinos from $K^+$ decay. The stark difference in energies of these two populations is driven by the off-axis angle of the detector and the difference in mass of the $\pi$ and $K$. The flux uncertainties used throughout this document are calculated based on SBN's EventWeight package~\cite{sbneventweight}, which in turn uses the original uncertainties developed by MiniBooNE~\cite{PhysRevD.79.072002}.

\begin{table}[!h]
  \centering
  \begin{tabular}{c|c|c|c|c}
    Sample & Off-Axis Angle (\unit{\degree}) & Baseline (\unit{km}) & $E^{\mathrm{Peak}}_{\nu}$ (\unit{GeV}) & $L/E^{\mathrm{Peak}}_{\nu}$ (\unit{km/GeV})\\
    \hline\hline
    NuMI        & 0.8  & 1.0 & 2.0 & 0.5  \\
    BNB ($K$)   & 9.2 & 0.77 & 1.4 & 0.55\\
    BNB ($\pi$) & 9.2 & 0.77 & 0.2 & 3.85\\ 
  \end{tabular}
  \caption{Comparison of off-axis angle, baseline, peak neutrino energy, and $L/E$ for the NuMI and BNB ($\pi/K$) samples.}
  \label{tab:bnb_numi_comparison}
\end{table}

\begin{figure}[!b]
  \centering
  \includegraphics[width=0.49\linewidth]{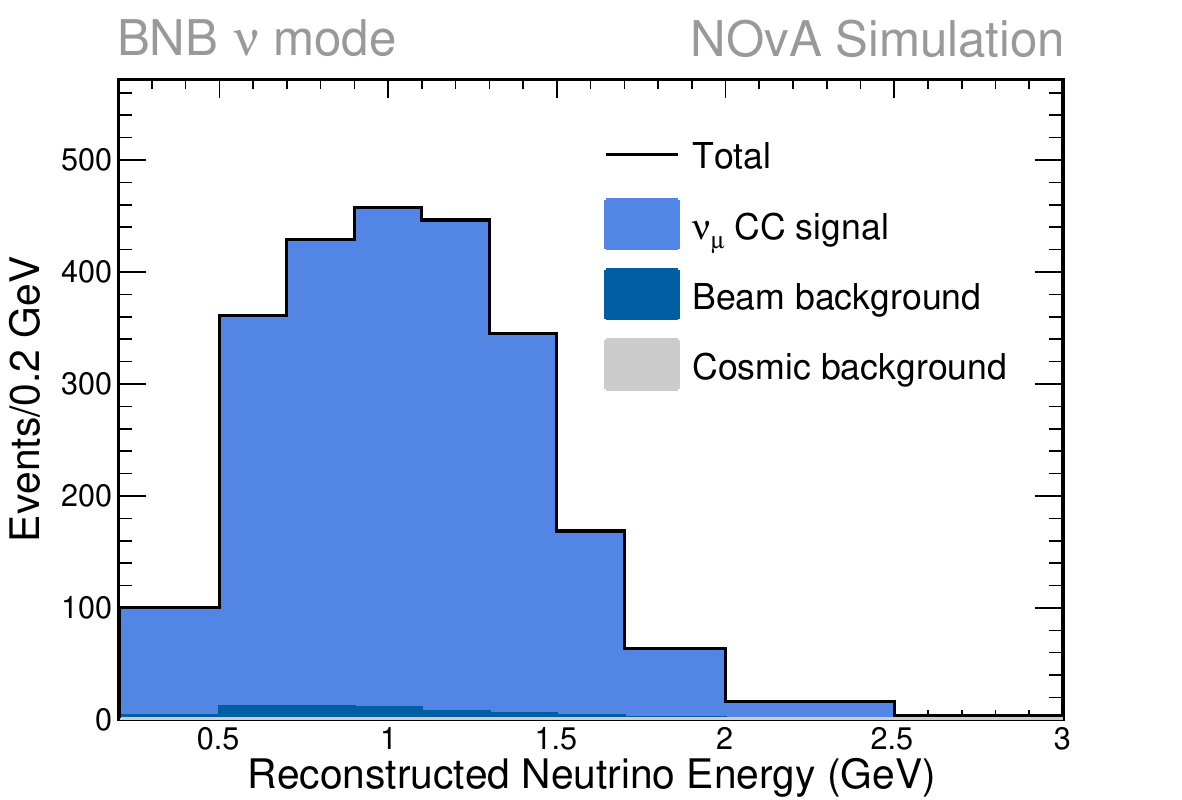}
  \includegraphics[width=0.49\linewidth]{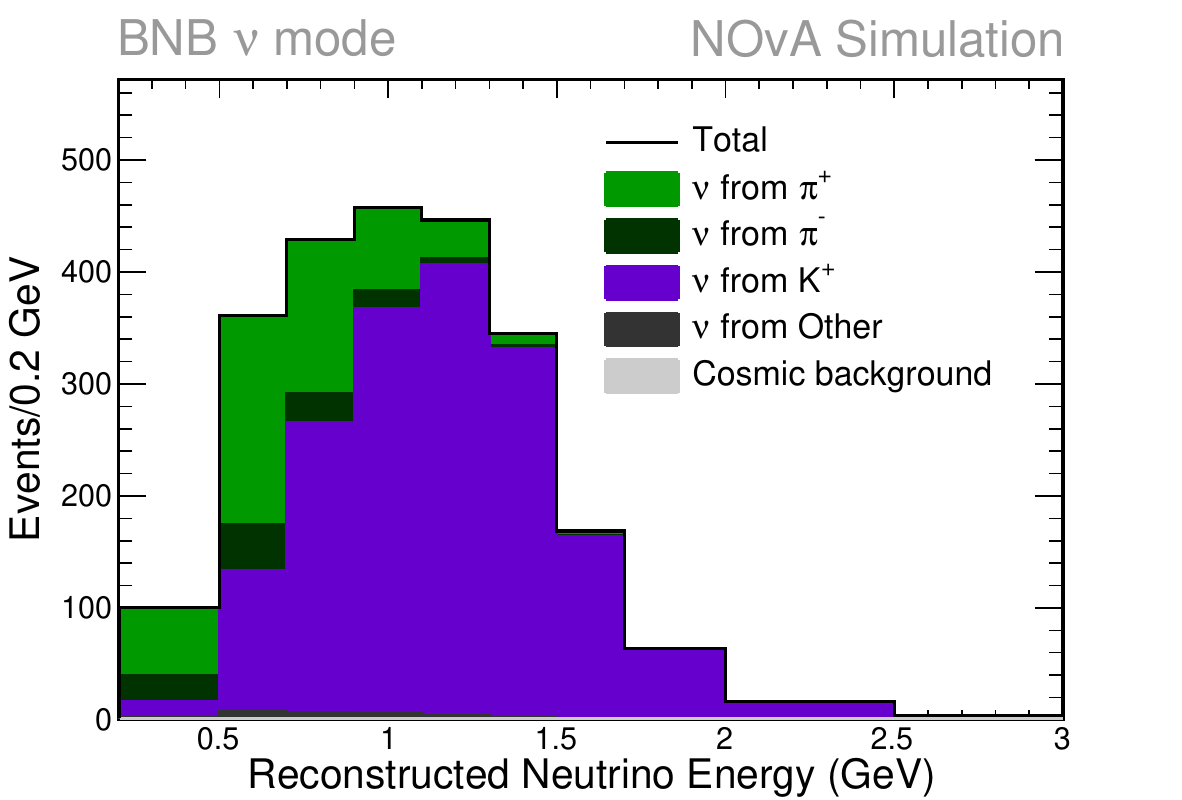}
  \includegraphics[width=0.49\linewidth]{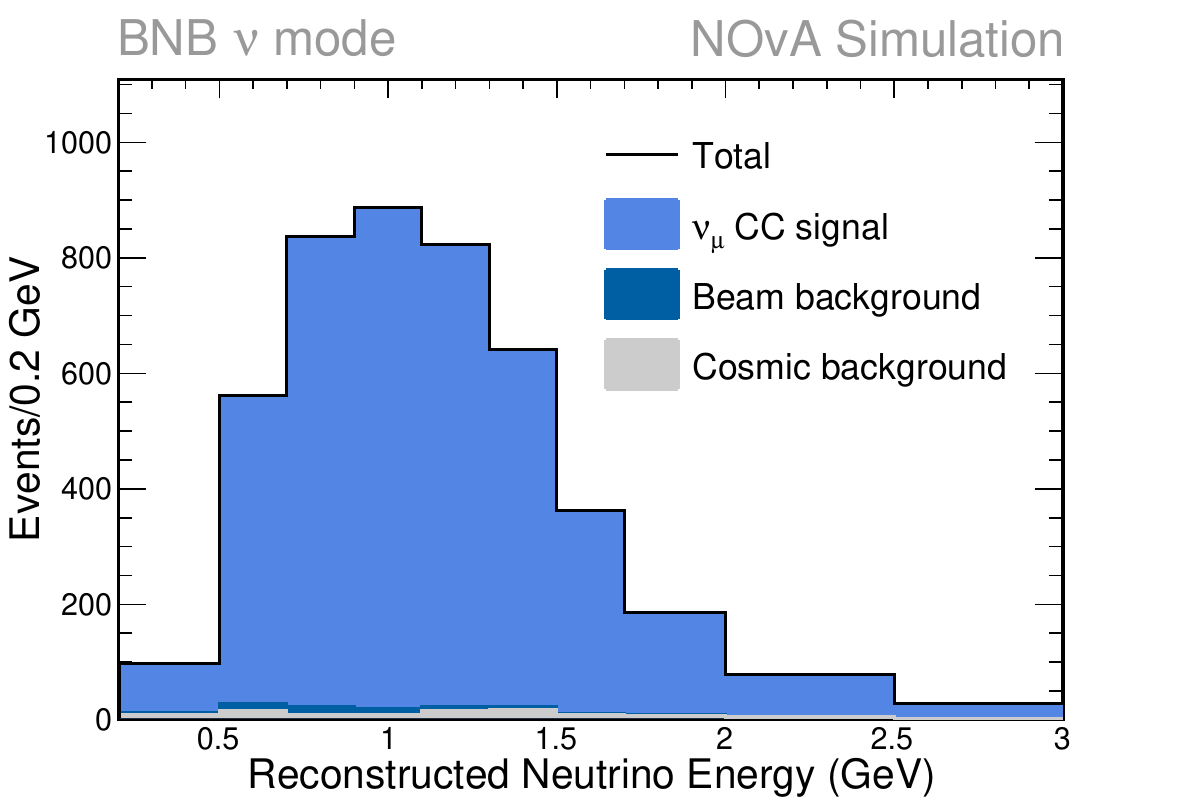}
  \includegraphics[width=0.49\linewidth]{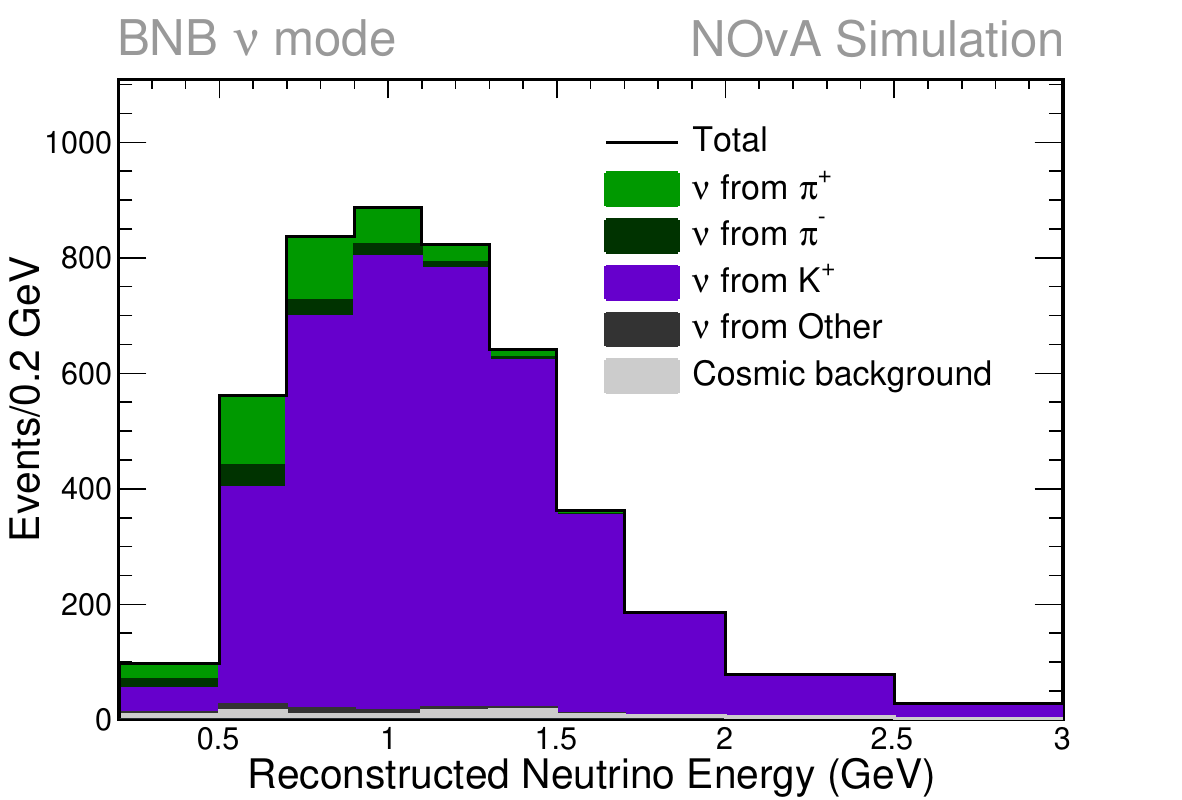}
  \caption{Distribution of fully contained events (top) and partially contained events (bottom) from a basic muon neutrino selection, shown broken out by signal/background (left) and by neutrino parent (right).}
  \label{fig:spectra}
\end{figure}

A summary of some of the differences between the NuMI and BNB samples are shown in Table~\ref{tab:bnb_numi_comparison}. The information in this table can be used to demonstrate why the inclusion of the BNB sample should improve NOvA's sterile neutrino search. The peak energy of the neutrinos from $K$ being around \qty{1.4}{GeV}, along with the \qty{770}{m} baseline produces an $L/E_{\nu}$ that is very similar to that of NuMI neutrinos, though at a different energy. In general, systematic uncertainties change as a function of $E_{\nu}$, but are independent of the baseline, while oscillations go as $L/E_{\nu}$. Probing the same $L/E_{\nu}$ at a different $E_{\nu}$ therefore gives us a handle to disambiguate systematic uncertainties from any potential sterile neutrino oscillations. An additional benefit is that the BNB decay pipe is \qty{50}{m}, around 6.5\% of the distance between the BNB target and the NOvA detector, whilst the NuMI decay pipe is \qty{675}{m}, around 67.5\% of the distance from the NuMI target to the ND, meaning oscillations in the BNB should be sharper and easier to identify.

With the flux simulation in-hand, and the needed modifications made to the NOvA software stack, we have produced a simulation of BNB neutrinos at the NOvA ND using NOvA's full simulation and reconstruction. With application of basic quality requirements, and using a convolutional neural network to select out muon neutrino interactions \cite{PhysRevD.100.073005}, we are able to select more than 5000 muon neutrino interactions in $2.5\times10^{21}$ POT, roughly equivalent to the data we had collected through the beginning of 2025, with a purity greater than 95\%. Neutrino energy spectra are shown in Fig.~\ref{fig:spectra} for fully contained and partially contained interactions. Around 35\% of selected interactions are fully contained, with the remaining interactions being partially contained. The distributions are split out by signal/background and by neutrino parent. The low-energy peak from neutrinos from pion decay is diminished in the selected event distributions due to the lower interaction cross section for these low-energy neutrinos, and because NOvA's reconstruction efficiency drops off significantly below 500 MeV. The lower peak energy than expected for neutrinos from kaons is due to the containment requirement for contained events, and due to escaping energy for partially contained events. 

\begin{figure}[!b]
  \centering
  \includegraphics[width=0.49\linewidth]{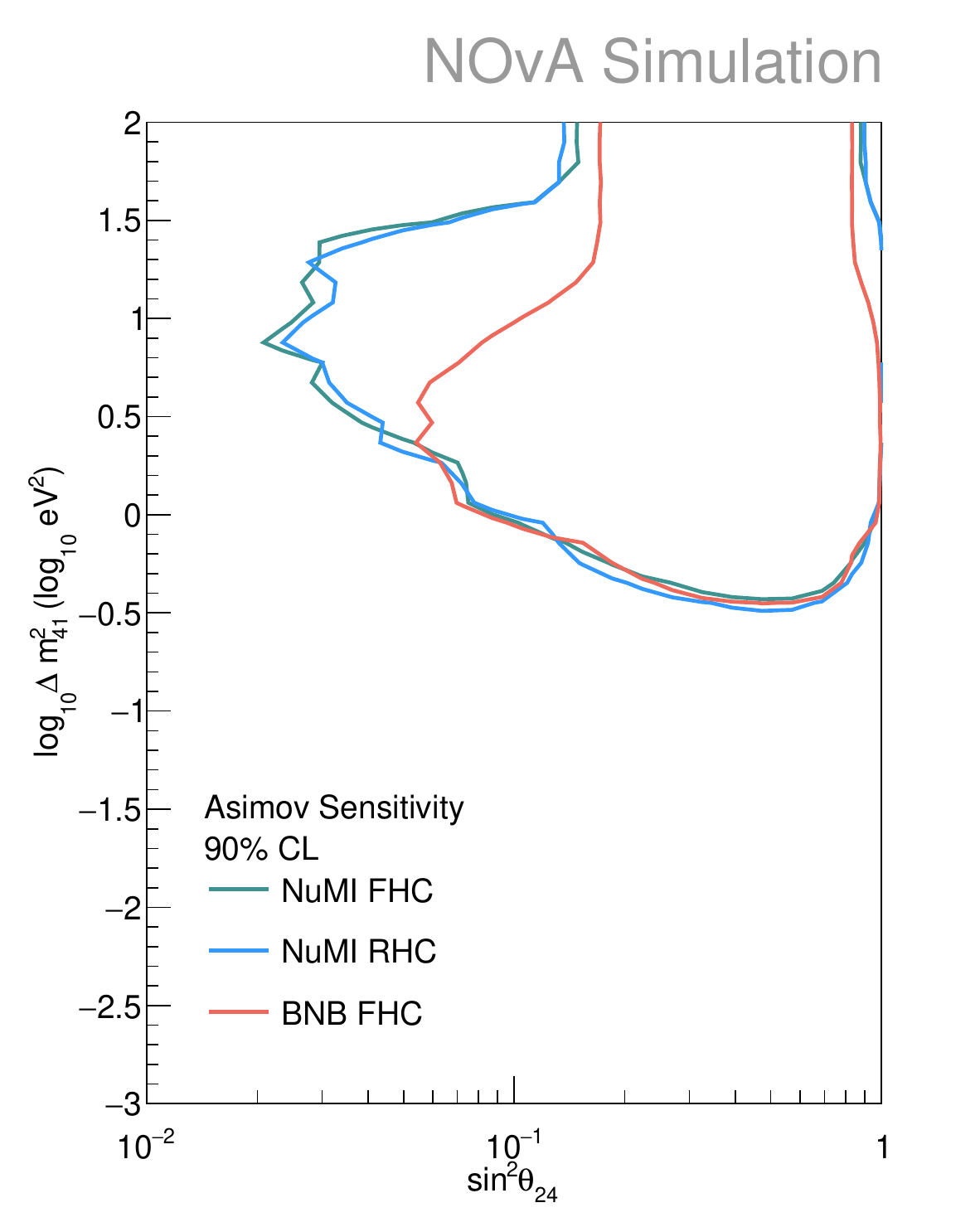}
  \includegraphics[width=0.49\linewidth]{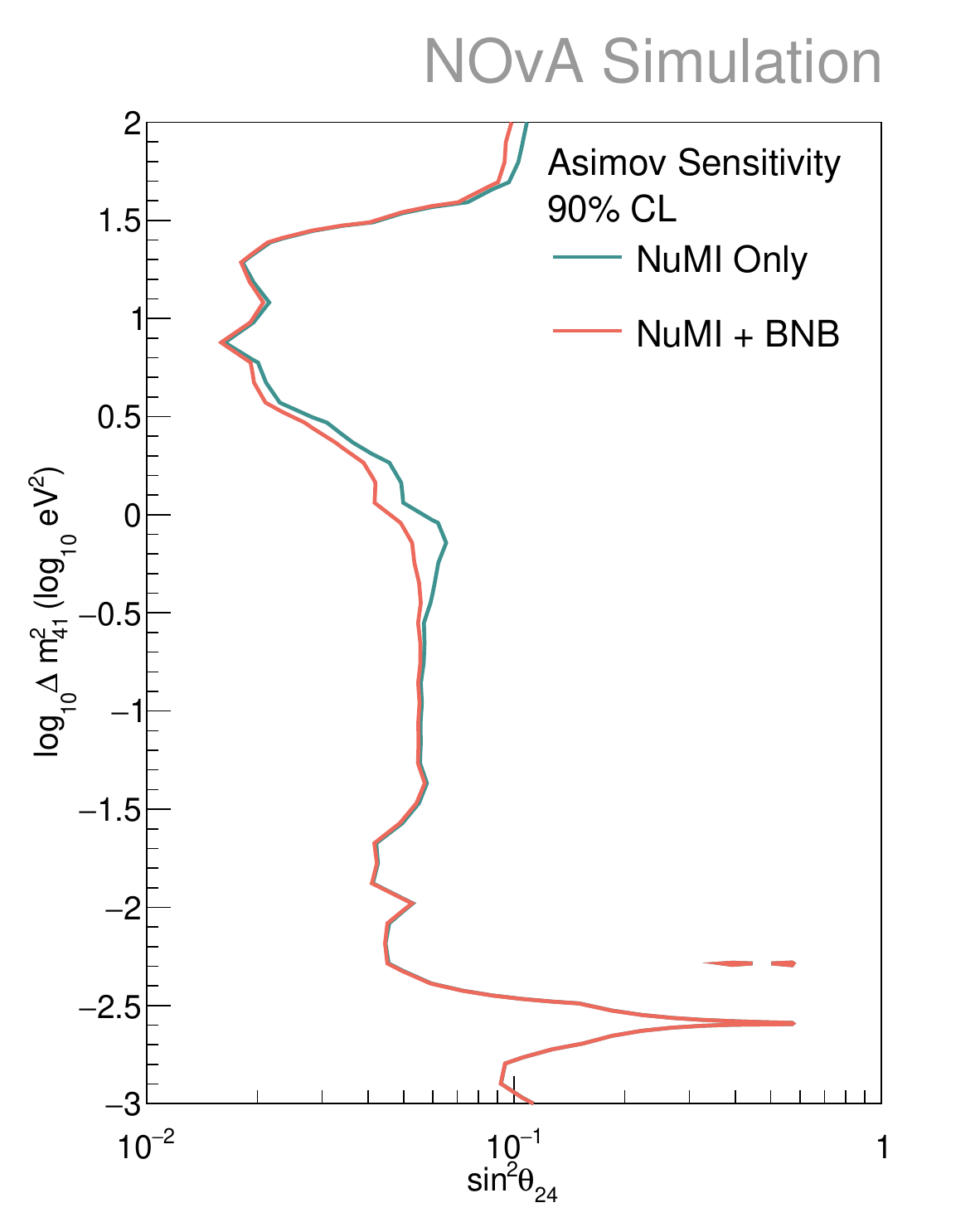}
  \caption{Asimov sensitivity plots using Wilk's theorem comparing the BNB dataset to the NuMI FHC and RHC samples (left), and comparing the all-sample sensitivity with and without the BNB sample (right). }
  \label{fig:sensitivities}
\end{figure}

Using these selected events, we are able to begin to understand the impact this dataset will have on NOvA's sterile neutrino search. Fig.~\ref{fig:sensitivities} (left) shows a comparison of the Asimov sensitivity from the BNB sample only with that of the NuMI FHC (neutrino-mode) and RHC (antineutrino-mode) samples, showing that for $\log_{10}\Delta m^2_{41} \lessapprox$ 0.5 $\log_{10}$eV$^2$, the BNB sample provides as much power as the other samples. As we move to larger values of $\Delta m^2_{41}$, the sensitivity for the BNB cuts in earlier than the NuMI samples, due to its relative weakness at higher energies, which is where most of the power in this region comes from. 

The sensitivity of the BNB dataset could be improved in several ways. The dataset has relatively low statistics compared to NuMI, meaning the statistical uncertainty is still relatively large. Additional data will therefore improve the sensitivity of this sample. Optimisation of the selection outlined here is underway, and could also potentially improve the sensitivity of the sample. Finally, improving NOvA's reconstruction at low energies could have a significant impact on the sensitivity. NOvA's current neural networks were designed to be performant at the higher energies of the NuMI beamline, and a dedicated re-training on lower energy events could improve our selection efficiency for neutrinos from $\pi$, around the 200 MeV peak.  

Fig.~\ref{fig:sensitivities} (right) shows a comparison of our all-sample Asimov sensitivity with and without the BNB sample. This initial study shows that the BNB sample increases our sensitivity by around 30\% for some regions of $\Delta m^2_{41}$, even without any of the planned improvements for this sample. 

\section{Conclusions and Outlook}

NOvA's most recent search for sterile neutrinos is world leading in a number of areas of parameter space. In the near future, we plan to update this analysis using almost twice the neutrino-mode dataset, our antineutrino dataset, and a sample of $\nu$-on-$e$ interactions, which will act as an in-situ flux constraint.

Looking into the future, we are investigating a number of novel techniques to further improve our sensitivity, from novel ways to constrain our uncertainties to new samples which can provide additional power to the search. In these proceedings, I have outlined one such sample we are in the process of incorporating, data from the BNB. Initial studies here show promise, with an improvement in our sensitivity of around 30\% beyond the upcoming analysis. We are currently investigating whether this sample can be made even more powerful through improved reconstruction and optimisation of the selection.

Beyond its use in NOvA's sterile neutrino search, the BNB sample at the NOvA ND could prove useful for experiments taking data on-axis in the BNB. The muon neutrino sample outlined in this work could be used to constrain the $K^+$ component of the on-axis BNB. This component produces a significant number of the intrinsic electron neutrinos in the beam, so a reduced uncertainty here could have a large impact on on-axis sterile neutrino searches and cross-section measurements.

\clearpage

\bibliographystyle{ieeetr}
\bibliography{bibliography}




\end{document}